\documentclass[preprint,aps,prc,showpacs,nofootinbib]{revtex4}
\usepackage{epsfig}
\usepackage{graphicx,amsmath}
\newcommand\ba{\begin{eqnarray}}
\newcommand\ea{\end{eqnarray}}
\newcommand\nn{\nonumber}
\newcommand{\be}{\begin{equation}}
\newcommand{\ee}{\end{equation}}
\newcommand\alb{\begin{align}}
\newcommand\ale{\end{align}}
\begin{document}
\title{Phenomenological analysis of near threshold periodic modulations of the proton time-like form factor} 

\author{A.~Bianconi}
\email[E-mail: ] { andrea.bianconi@unibs.it}
\affiliation{Dipartimento di Ingegneria dell'Informazione, 
Universit\`a di Brescia \\ and 
Istituto Nazionale di Fisica Nucleare, Gruppo Collegato di Brescia, 
25133 Brescia, Italy
}
\author{E.~Tomasi-Gustafsson}
\email[E-mail: ]{etomasi@cea.fr}
\affiliation{CEA,IRFU,SPhN, Saclay, 91191 Gif-sur-Yvette Cedex, France}

\begin{abstract}

We have recently highlighted the presence of a periodically  
oscillating 10 \% modulation in the BABAR data on the proton time-like form 
factors, expressing the deviations from 
the point-like behavior of the proton-antiproton electromagnetic 
current in the reaction 
$e^++e^-$ $\rightarrow$ $\bar{p}+p$. Here we deepen our previous data analysis, and confirm 
that in the case 
of several standard parametrizations it is possible to write the 
form factor in the form $F_0$ $+$ $F_{osc}$, where $F_0$ is a 
parametrization expressing the long-range trend of the form 
factor (for $q^2$ ranging from the $\bar{p}p$ threshold to 
36 GeV$^2$), and $F_{osc}$ is a function of the form $\exp(-Bp)\cos(Cp)$, 
where $p$ is the relative momentum of the final $\bar{p}p$ pair. Error 
bars allow for a clean identification of the main features of this  
modulation for $q^2$ $<$ 10 GeV$^2$. Assuming this oscillatory modulation to be an effect of final state 
interactions between the forming proton and the antiproton, we propose 
a phenomenological model based on a double-layer imaginary optical 
potential. This potential is flux-absorbing when the distance  
between the proton and antiproton centers of mass is $\gtrsim$ 1.7-1.8 fm 
and flux-generating when it is $\lesssim$ 1.7-1.8 fm. The main features of 
the oscillations may be reproduced with some freedom in the potential 
parameters, but the transition between the two layers must be sudden 
(0-0.2 fm) to get the correct oscillation period. The flux-absorbing part of the $\bar{p}p$ interaction is well known 
in the phenomenology of small-energy antiproton interactions, and is due 
to the annihilation of $\bar{p}p$ pairs into multi-meson states. We interpret 
the flux-creating part of the potential as due to the 
creation of  a $1/q$-ranged state when the virtual photon decays into a set of 
current quarks and antiquarks. This short-lived compact state may be 
expressed as a sum of several hadronic states including the ones with large 
mass $Q_n\gg q$, that may exist for a time $t\sim 1/(Q_n-q)$. 
The decay of these large mass states leads to an intermediate stage 
regeneration of the $\bar{p}p$ channel. 

\end{abstract}

 \pacs{
      {12.40.Nn} {Regge theory, duality, absorptive/optical models}
      {13.40.Gp} {Electromagnetic form factors }
} 
\maketitle
\section{Introduction}

Both the annihilation reactions 
\ba
&&e^+ + e^- \to \bar p +p ,
\label{eq:eq1}\\
&&\bar p + p \to e^+ + e^- ,
\label{eq:eq2}
\ea
have been used to extract the electromagnetic form factors (FFs)
of the proton in the time-like (TL) region (for a recent review see \cite{Pacetti:2015iqa} and references therein). Assuming that the interaction occurs through one photon exchange, the annihilation cross section is expressed in terms of the FF moduli squared, as  FFs are of complex nature  in the explored  kinematical region \cite{Zichichi:1962ni}. 

The collected statistics 
has not permitted the individual determination  of the electric ($G_E$) and  magnetic ($G_M$) FFs 
due to the available limited luminosity.  
The cross section $\sigma$ of the reactions (\ref{eq:eq1}) and (\ref{eq:eq2}), 
allows to extract the squared modulus of an effective 
form factor $F_p$, that is in practice equivalent to the assumption $G_E=G_M$ (strictly valid only at threshold) \cite{Bardin:1994am}:
\be
|F_p|^2=\displaystyle\frac{3\beta q^2 \sigma}
{2\pi\alpha^2 \left(2+\displaystyle\frac{1}{\tau}\right)}, 
\label{eq:Fp}
\ee
where $\alpha=e^2/(4\pi)$, $\beta=\sqrt{1-1/\tau}$, 
$~\tau={q^2}/(4M^2)$, $q^2$ is the squared invariant mass of the colliding 
pair, and $M$ is the proton mass. The effect of the Coulomb singularity 
of the cross section at the $\bar{p}p$ 
threshold is removed by the $\beta$ factor: 
$\beta$ $\rightarrow$ 0 for $q$ $\rightarrow$ $2M$, so that $\beta \sigma$ 
is finite and the effective form factor is expected to be finite at the 
threshold. 

The 
reactions  (\ref{eq:eq1}) and (\ref{eq:eq2}) test 
close-distance components of the wave function of the 
$\bar{p}p$ system, that are supposed to be suppressed because of 
$\bar{p}p$ annihilation. 
Data on the $\bar{p}$-nucleon and $\bar{p}-$nucleus 
annihilation process at low energies 
(see \cite{Balestra:1985kn,Balestra:1989rb,Bizzarri:1973sp,Bruckner:1989ew,Balestra:1984wd,Balestra:1985wk,Bertin:1996kw,Benedettini:1997fk,Zenoni:1999su,Zenoni:1999st,Bianconi:2000nh,Bianconi:1999vq,Bianconi:2011zz},  
and the related theoretical 
analyses \cite{Bianconi:2000ap,Batty:2000vr,Friedman:2014vva})  
show that a proton 
and an antinucleon overlap little. When their surfaces come in touch, or even 
within a distance of 1 fm, they 
annihilate into other hadron states. Elastic scattering is present, but 
either of diffractive origin (for $p\gg 100$~MeV), or of refractive 
repulsive hard-core nature (near threshold). In all 
cases, the wave function of the $\bar{p}p$ relative motion is 
estimated to be strongly suppressed at distances lower than 1 fm. 
On the other side, reactions (\ref{eq:eq1}) and (\ref{eq:eq2}) 
involve a virtual photon with 
center of mass (c.m.) energy  $\sqrt{s}\geq 2 M\approx $ 2 GeV. This 
means that the 
space-time regions where the virtual photon is formed (or decays) have size 
$\Delta r \leq 0.1$ fm. So, the virtual photon tests the 
short-distance components of the $\bar{p}p$ system, and TL FF 
complement, in this respect, the information acquired from other annihilation 
experiments. 

Until recently, uncertainties and discontinuities between 
data coming 
from different measurements have prevented from appreciating the continuity 
features of TL FF data over a large $q^2$ range. 
The recent  results from the BABAR 
collaboration \cite{Lees:2013xe,Lees:2013uta}, cover a $q^2$ range 
going from near the threshold to 36 GeV$^2$, with more than 30 data 
points only in the region $q^2$ $<$ 10 GeV$^2$. 

Specific features of the effective FF related to final state 
interactions between $\bar{p}$ and $p$ appear when  
expressing $|F_p(q^2)|^2$ in terms of the 3-momentum of the relative 
motion of the two hadrons. 
This has been illustrated in a recent 
work \cite{Bianconi:2015owa}, where we have highlighted periodic 
features in a modulation of the order of 10 \% , superimposed on the long-range trend of the 
effective FF. 
The precision of the available data does not forbid the 
interpretation where the oscillation 
pattern is attributed to independent   
resonant structures, as in Ref.  \cite{PhysRevD.92.034018}. 
However, the underlying assumption of the present work is that the oscillations are expression of a unique 
interference mechanism, affecting all the $q^2$ 
range where the oscillations are visible. Of course, the two points 
of view may coexist within a model where two or more resonance poles are 
the result of a global underlying mechanism, as in \cite{Brodsky:2007hb},
or within a duality framework. As observed in our final discussion, the proposed 
phenomenological interpretation may be framed  
within several models, including multiple-pole ones. 

In the present work we first scrutinize these oscillations by expanding  
the data analysis of Ref. \cite{Bianconi:2015owa}. We use four different 
parametrizations from the literature for the so-called "background'' 
term (the effective form factor as it appears if one neglects the small 
oscillating modulation). For each choice of the background, we fit the 
residual modulation, visible in the difference between the data and the background fit. 
We analyze the uncertainty on the periodical character of the 
oscillations, and on their possible long-range scaling behavior (Section II). 

Next, we present a phenomenological model for the rescattering origin of the 
oscillations, within a DWIA (distorted wave impulse approximation) 
scheme where the outgoing (or incoming) hadron waves are distorted 
by an optical potential (Section III). In absence of distorting potentials, the background 
form of the TL FF is recovered (Section IV). This analysis shows that it is possible 
to reproduce  most of the features of the observed oscillations this way, 
but important constraints must be satisfied by the rescattering potential (Section V).
Conclusions summarize the main finding of the paper. 

\section{Analysis of the data}

\begin{figure}
\mbox{\epsfxsize=10.cm\leavevmode \epsffile{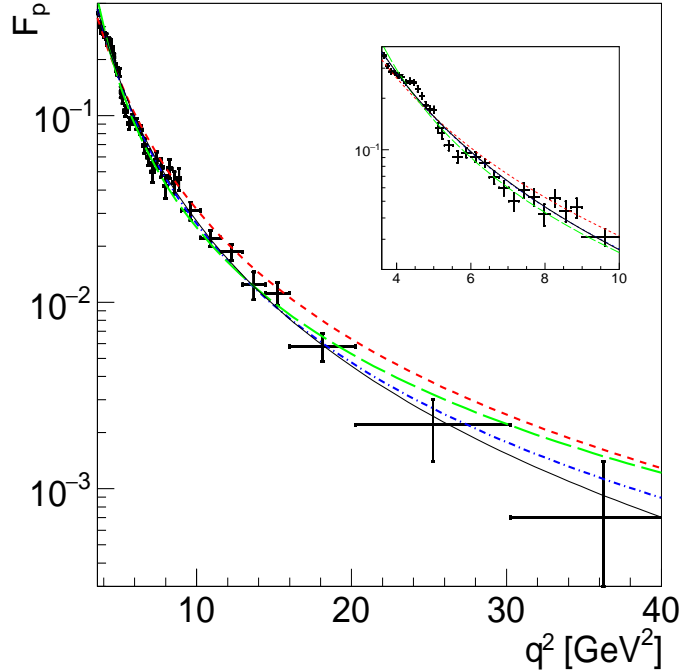}}
\caption{(color online) Data on the TL proton generalized FF as a function of $q^2$, from Ref. \cite{Lees:2013xe,Lees:2013uta}
together with the fits from 
Eq. (\ref{eq:mpr}) (black solid line), 
Eq. (\ref{eq:qcd}) (blue dash-dotted line), 
Eq. (\ref{eq:eak}) (red dashed line), and 
Eq. (\ref{eq:BdT}) (Green, long-dashed line).
The insert 
magnifies the near threshold region. Because of their large 
error bars, the points over 16 GeV$^2$ do not affect fit parameters, so that 
the four fits best reproduce the data in the insert, apart for the 
oscillations that are the focus of this work. 
}
\label{Fig:WorldData}
\end{figure}

The effective proton  FF extracted from BABAR data on  $e^++e^-$ 
$\rightarrow$ $\bar{p}+p (\gamma)$  \cite{Lees:2013xe,Lees:2013uta}, is 
reported 
in Fig. \ref{Fig:WorldData}  (black circles) as a function of $q^2$, 
that is equivalent to the total energy squared $s$ in the TL region. 
As it can be noticed in the insert that highlights the near threshold 
region, $4M^2\le q^2\le 10 $ GeV$^2$,  the data show irregularities. 
These irregularities acquire a peculiar structure 
when $q^2$ is replaced by the relative momentum 
of the $\bar p p$ system in the rest frame of one of the hadrons \cite{Bianconi:2015owa}. 

We introduce a function of the form 
\be
F(p)\ \equiv\ F_0(p)\ +\ F_{osc}(p)
\label{eq:fit1}
\ee
 where 
\begin{enumerate}

\item The 3-momentum 
\be
p(q^2)\ \equiv\ \sqrt{E^2-M^2},\hspace{0.3truecm} E\ \equiv\ q^2/(2M)\ -\ M. 
\label{eq:plab}
\ee
is the momentum of one of the two hadrons in the frame where the other one 
is at rest. 

\item  $F_0(p)$  ("regular background 
term" ) is a function expressing the regular 
behavior of the FF over a long $q^2$ range. 

\item  $F_{osc}(p)$ describes the deviation of the TL FF from the long-range 
regular background appearing in the region $0\leq ~p ~\lesssim ~ 3$ GeV and  
corresponding to 
$M\leq ~ q ~\lesssim ~3$ GeV, with $q=\sqrt{q^2}$.

\end{enumerate} 

Different forms available 
from the literature can be used for the background term. 
As measured by BABAR, 
$F_0[p(q)]$ is slightly steeper than expected on the 
ground of the corresponding fits of the space-like (SL) FF (dipole-like shape) 
and of the power law corresponding to quark-counting 
rules \cite{Matveev:1973uz,Brodsky:1973kr}. 
The recent data are best reproduced by the  function $F_R$ 
proposed in \cite{TomasiGustafsson:2001za} that we will use as a reference: 
\begin{align}
&|F_R|(q^2)|\ =\ 
\frac{\cal A}{(1+q^2/m_a^2)\left[1-q^2/0.71 \right]^2},
\nonumber
\\
&{\cal A}=7.7~\mbox{GeV}^{-4},
\ m_a^2=14.8~\mbox{GeV}^2.
\label{eq:mpr}
\end{align}

Other parametrizations have been proposed. The world data prior to BABAR 
results were well 
reproduced in the experimental papers \cite{Ambrogiani:1999bh} according 
to the function: 
\be
|F_S(q^2)|=\displaystyle\frac{\cal A}{(q^2)^2\log^2(q^2/\Lambda^2)}, 
\label{eq:qcd}
\ee
where $q^2$ is expressed in GeV$^2$, ${\cal A}=40$ GeV$^{-4}$ and 
$\Lambda=0.45$ GeV$^{2}$. 

The functional form of Eq. (\ref{eq:qcd}) is driven by the extension to 
the TL region of the dipole behavior. The dipole model of the SL 
FFs, more precisely their $(q^2)^2$ dependence, is empirically well known 
since the first elastic scattering experiments \cite{Hofstadter:1960zz} 
and agrees with most of the nucleon models developed during last century, 
as for example, the constituent quark model of Ref. \cite{Yamada:1971ta}. 
It is also consistent with  
PQCD large-$q^2$ predictions \cite{Brodsky:1973kr}. 

Based on  Ref. \cite{Shirkov:1997wi}, in order 
to avoid ghost poles in $\alpha_s$, the following modification was suggested (\cite{Kuraev}) :
\be
|F_{SC}(q^2)|\ =\ 
\displaystyle\frac{\cal A}{(q^2)^2\left [ \log^2(q^2/\Lambda^2)+\pi^2 \right ] }. 
\label{eq:eak}
\ee
In this case the best fit parameters are ${\cal A}=72$ GeV$^{-4}$ 
and  $\Lambda=0.52$ GeV$^{2}$. 

In Ref. \cite{Brodsky:2007hb} a form was suggested with two poles of 
dynamical origin (induced by a dressed electromagnetic current) 
\be
|F_{TP}(q^2)|\ =\ 
\displaystyle\frac{\cal A}{(1-q^2/m_1^2) (2-q^2/m_2^2) }. 
\label{eq:BdT}
\ee
The best fit parameters are ${\cal A}=1.56$,  $m_1^2=1.5$ GeV$^{2}$ and $m_2^2=0.77 $ GeV$^{2}$. The parametrizations with the best fit parameters are illustrated in Fig. \ref{Fig:WorldData} and  summarized in  Table \ref{tab:ffit}. 
\begin{figure}
\mbox{\includegraphics[width=9.cm]{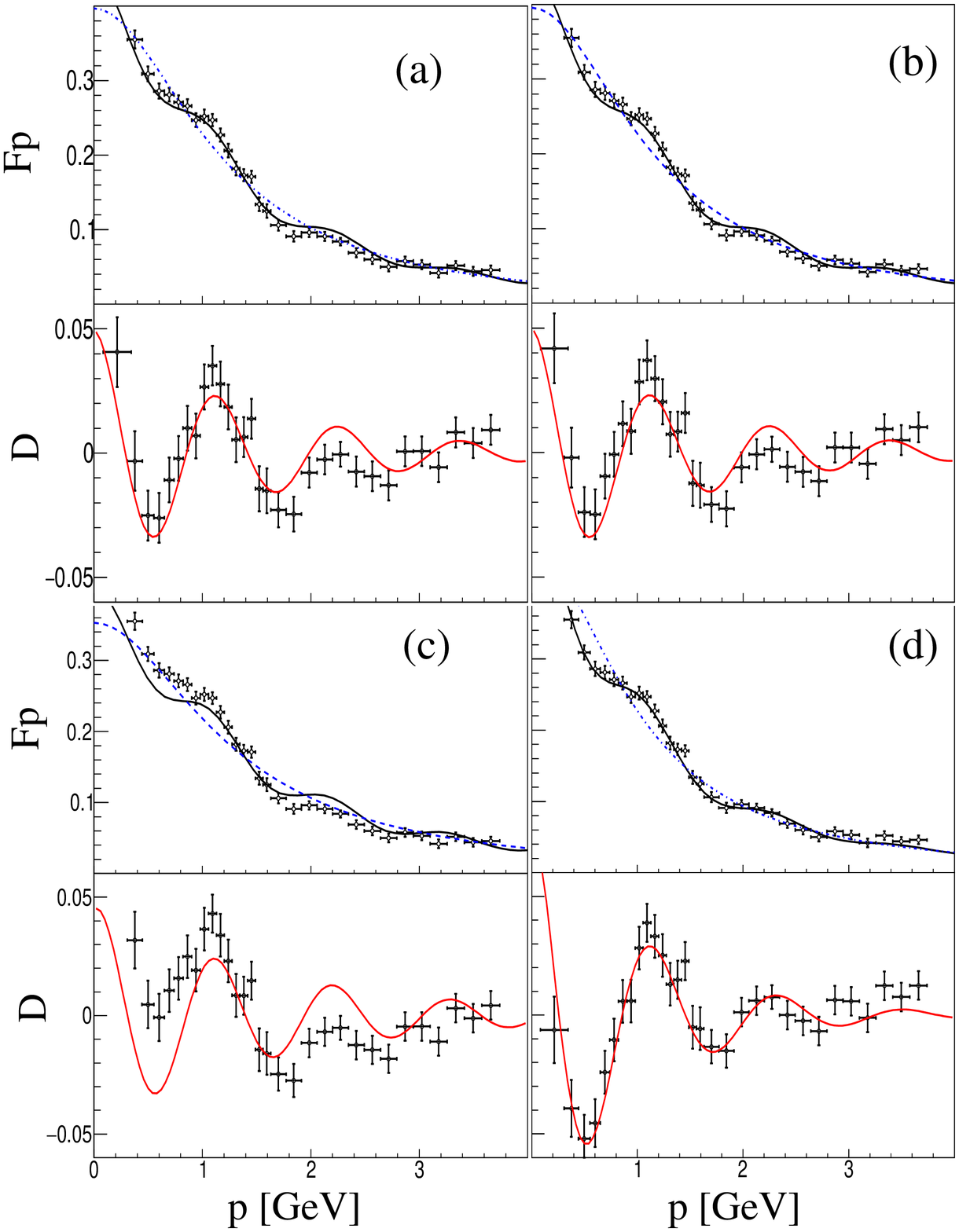}}
\caption{Referring to Eqs. (\ref{eq:fit1}) and (\ref{eq:plab}), we report the 
background fits $F_0(p)$ 
of the BABAR data, according to the four parametrizations 
a) $F_0 = F_R$ from \cite{TomasiGustafsson:2001za} (Eq. (\ref{eq:mpr}), see text 
or Table 1), 
b) $F_0 = F_S$ from \cite{Ambrogiani:1999bh} (Eq. (\ref{eq:qcd})), 
c) $F_0 = F_{SC} $ from \cite{Shirkov:1997wi} (Eq. (\ref{eq:eak})), 
d) $F_0 = F_{TP}$ from \cite{Brodsky:2007hb} (Eq. (\ref{eq:BdT})), and the 
corresponding fits $F_{osc}(p)$ of the differences between the data and each 
parametrization. In all the four cases $F_{osc}$ has the damped 
oscillation form of Eq. (\ref{eq:diff_fit}), with the best-fit parameters 
reported in Table 1. 
For each insert: (top) the data of BABAR are plotted, together with 
the parametrization $F_0(p)$ (blue, dashed line) and the complete fit 
$F_R(p)=F_0(p) + F_{osc}(p)$ (solid black line); 
(bottom) the difference of the data and the parametrization is shown, 
together with the fit $F_{osc}(p)$ (solid red line). }
\label{Fig:allfit}
\end{figure}
\begin{table*}[h]
\caption{Background fit functions from Eqs. (\ref{eq:mpr},
\ref{eq:qcd}, \ref{eq:eak}, \ref{eq:BdT}) 
(see Fig. {Fig:WorldData}), 
and parameters $A$, $B$, $C$, $D$ (with the related $\chi^2$/n.d.f.) 
for Eq. (\ref{eq:diff_fit} )
fitting in each case the difference between the data and the corresponding 
background function.}
\label{tab:ffit}
\centering\small
\begin{tabular} {l c lcl cl clc lc ll}
\hline\hline
Ref. &Background function  
                                       & \emph{A$\pm \Delta $A}
                                       & \emph{B$\pm \Delta $B}
                                       & \emph{C$\pm \Delta $C}
                                       & \emph{D$\pm \Delta $D}
                                       & $\chi^2$/n.d.f.
                                       \\
&\textrm{\emph{}}
                                       &    
                                       & [GeV]$^{-1}$
                                       & [GeV]$^{-1}$ & 
                                       & \emph{}
                                       & &
                                      \\
\hline\hline
\cite{TomasiGustafsson:2001za} &$|F_R|=\displaystyle\frac{\cal A}{(1+q^2/m_a^2)\left[1-q^2/0.71 \right]^2}$ & 0.05$\pm $0.01  & 0.7$\pm $0.2 & 5.5$\pm$0.2   &0.0$\pm$0.3 & \ \ \ 1.4\\
 &$ {\cal A}=7.7~\mbox{GeV}^{-4},
\ m_a^2=14.8~\mbox{GeV}^2 $  &   &  &    & & \\
\hline
\cite{Ambrogiani:1999bh} &$|F_S|=\displaystyle\frac{\cal A}{(q^2)^2\log^2(q^2/\Lambda^2)}$
                            & 0.05$\pm $0.01  & 0.7$\pm $0.2 & 5.5$\pm $0.2   &0.1$\pm $0.3 &\ \ \ 1.3\\
 &${\cal A}=40$ GeV$^{-4}$,  $\Lambda=.45$ GeV$^{2}$ 
                            &   &  &  & & \\
\hline
\cite{Shirkov:1997wi} &$|F_{SC} | = \displaystyle\frac{\cal A}{(q^2)^2\left [ \log^2(q^2/\Lambda^2)+\pi^2 \right ] }$   & 0.05$\pm $0.01 & 0.6$\pm $0.2 & 5.8$\pm $0.2  &0.1$\pm $0.3 &\ \ \ 4.0\\
&${\cal A}=72$ GeV$^{-4}$, $\Lambda=0.52$ GeV$^{2}$  
 &   &  &  & & \\
\hline
\cite{Brodsky:2007hb} &$|F_{TP}|= 
\displaystyle\frac{\cal A}{(1-q^2/m_1^2) (2-q^2/m_2^2) }$
                              & 0.1$\pm $0.01  & 1.0$\pm $0.2 & 5.3$\pm $0.2   &0.2$\pm $0.3  & \ \ \ 1.0\\
 &${\cal A}$=1.56,  $m_{1,2}^2=1.5, 0.77$ GeV$^{2}$
                            &   &  &  & & \\
\hline\hline
\end{tabular}
\end{table*}
The best fit functions are then subtracted from the data, leaving a regular oscillatory behavior,  Fig. \ref{Fig:allfit}.  It has magnitude 
$\sim$ 10 \% of the regular term, and is well visible over the data 
uncertainties for $p$ $>$ 3 GeV. We have fitted the 
difference between the BABAR data and the regular background term $F_0(p)$ 
with the 4-parameter function 
\be
F_{osc}(p)\ \equiv\ A\ \exp(- Bp)\ \cos(C p + D).
\label{eq:diff_fit}
\ee
Let us focus on the case $F_0$ $=$ $F_R$, Eq. (\ref{eq:mpr}). The corresponding 
difference data are plotted in the lower panel of Fig. \ref{Fig:allfit}a. 
The relative errors in the parameters $C$ and $B$ show that the oscillation period is better defined than the 
damping coefficient $\exp(-Bp)$. 
Two and a half oscillations are clearly visible over the reaction threshold,  
while for $p>$ 2.8 GeV the vertical error bars overcome the oscillation 
amplitude $A ~\exp(-Bp)$. 

The parameter $D$ defines the position of the first oscillation maximum that 
occurs at $p=0$ within the error $\Delta D~ P /(2\pi)$, where $P$ is the oscillation 
period. Estimating 
the oscillation period $P=$ 1.13 GeV, the first oscillation maximum 
occurs at $p=0$ within an error of 0.05 GeV. 
Five  peaks (maxima and minima)  are visible and the periodicity hypothesis, that is implicit in the $\cos(Cp+D)$ term implies that they are regularly spaced by a half-period of 1.13/2 GeV. 
Examining Fig. \ref{Fig:allfit}a (lower panel) we find: 
\begin{itemize}
\item 1st maximum: $p$ $=$ $0 \pm 0.05$ GeV (from the fit error),

\item 1st minimum: estimated at 0.57 GeV, visible inside the range 0.5-0.6 GeV, 

\item 2nd maximum: estimated at 1.13 visible at 1.1 GeV with negligible uncertainty,

\item 2nd minimum: estimated at 1.7 GeV, visible inside the range 1.7-1.8 GeV,

\item 3rd maximum: estimated at 2.26 GeV, visible inside the range 2.2-2.3 GeV,

\item 3rd minimum: estimated at 2.83 GeV, visible inside the range 2.6-2.9 GeV. 

\end{itemize}
The largest uncertainty in the position of the 
peak is found in the last case. Excluding this one, 
in the other cases the relative discrepancy lies within (0.1 GeV / 0.57 GeV) 
$\approx$ 15 \%. This justifies the presence of the periodic 
term $\cos(Cp+D)$ in the fit. Such analysis may be repeated for the other 
cases in Fig. \ref{Fig:allfit}, with similar results.

Concerning the 
amplitudes of the half-oscillations, each of them is about $1/\sqrt{2}$ of 
the previous one, so that each maximum of $F_{osc}(p)$ 
is about 1/2 of the previous maximum. This motivates the use of 
$\exp(-Bp)$ in the fit, although in this case the error on this 
``$1/\sqrt{2}$ rule'' is too large to exclude good fits with other 
analytic shapes. In particular, one possibility is that the oscillation 
amplitude is proportional to the background term, so that the 
overall fit would assume the form $F_0(p)[1 + \epsilon \cos(Cp)]$ 
with constant $\epsilon$ $\approx$ 0.1. 
For increasing momenta within the visible range, the damping 
of the oscillation and of the regular background term are similar: 
$F_{osc}(p)/F(p) \approx$ constant, both decreasing by $1/e$ in about 
1.4-1.5 GeV. 

Since increasing relative errors hide the possible presence 
of the oscillations for $p$ $>$ 3 GeV, we 
cannot know whether the oscillations are   
relevant at larger $p$ or not. 
Assuming that they are, the point of view supported in our previous 
work \cite{Bianconi:2015owa} and in the following is 
that we are facing 
an interference effect between a small number of  
amplitudes, effectively competing in the visible momentum range. 
These amplitudes must be few, not 
forming a regular continuum, otherwise they would give rise to a diffraction 
pattern, rather than an oscillation pattern. 

Note that the oscillatory behavior is present already  in other 
invariant functions of $q^2$,  but  not periodic. The relevant point is that $F_{osc}(p)$ is periodic with respect to $p$, 
not with respect to $q^2$ or $q$. 
Since $p$ is a variable that is uniquely associated with the relative 
motion of the hadron, we associate periodicity  
with interactions between the forming hadrons after the virtual 
photon has been converted into quarks and antiquarks, 
Eq. (\ref{eq:eq1}), or before quarks and antiquarks 
annihilate into a virtual photon, Eq. (\ref{eq:eq2}). 
In both cases we name "rescattering'' these interactions.

%
\section{Optical model}

We assume that rescattering is a relatively $small$ perturbation, and that  
in absence of rescattering the effective FF would coincide exactly with $F_0(p)$. We 
also assume that it is possible to neglect the dependence of 
the rescattering mechanism on $q^2$. 

Let $\vec r$ be the space variable that is Fourier-conjugated to $\vec p$: 
$r$ is the distance between the centers of mass of the two 
forming hadrons, in the frame where one is at rest. 
The observed behavior is modeled via a two-stage process where: 
\begin{itemize}
\item  In the $e^+e^-$ $\rightarrow$ $\bar{p}p$ "bare" process a 
$\bar{p}p$ pair is formed at a distance $r$ with space distribution 
amplitude $M_0(r)$. 
\item  Rescattering takes place between the newly formed hadrons ($p$ and $\bar{p}$) according to 
an optical potential that is function of their distance $r$. 
\end{itemize}

To introduce rescattering we use the Distorted Wave 
Impulse Approximation (DWIA) formalism, following the scheme employed in 
Ref. \cite{Bianconi:1996ur}. 
The starting point is the Fourier transform 
\ba
F_0(p)\ \equiv\ \int d^3 \vec r\exp(i \vec p \cdot \vec r) M_0(r)
\label{eq:c0}
\ea
where we interpret $\exp(i \vec p \cdot \vec r)$ as the plane wave 
final state of the $\bar{p}p$ pair in their center of mass, and 
$M_0(r)$ as a matrix element describing the earlier stage of the 
process. Neglecting rescattering, a detailed model for the formation process 
would lead to a matrix element of the form 
\ba
F_0(p) &=& <\psi_f(x_1,...,x_n)\psi_f(\vec r) |T(r,x_1,..,x_n,x_{e^+e^-})| 
\psi_i(x_{e^+e^-})>\nn \\
&\equiv&\ \int d^3 \vec r\ \psi_f(\vec r)\ M_0(r),
\ea
where the hard operator $T$ is sandwiched between 
the initial state $\psi_{i}$, that is function of the 4-vector 
$x_{e^+e^-}$ expressing the relative coordinates of the $e^+e^-$ pair, 
and the final states $\psi_f$ that depend on the internal 
coordinates $x_1, x_2, ...x_n$ of the two hadrons as well as on their 
relative position $\vec r$. 
The result of integrating over all variables 
but $r$, is $M_0(q^2,r)$, that in general depends on $q^2$ since this is 
a parameter of $\psi_{i}$ and $\psi_f$. We assume that this dependence 
may be neglected in the 
range $0 <p < 3$~GeV where the oscillations are distinguishable 
from the background. 
~
\begin{figure}
\mbox{\includegraphics[width=14.cm]{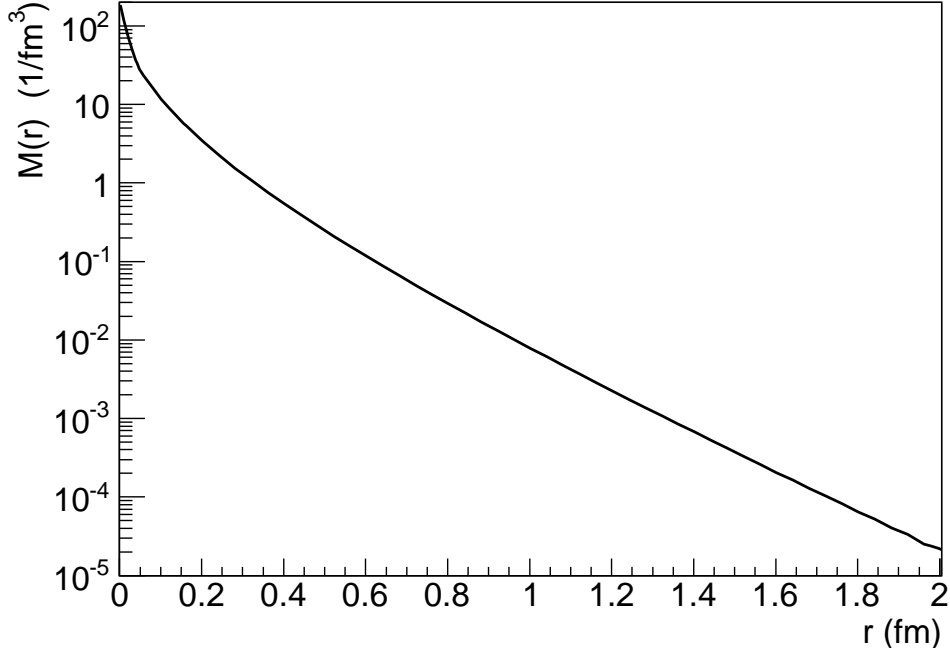}}
\caption{ 
The 3-dim Fourier transform $M_0(r)$ of $F_0(p)$, defined 
in Eq. (\ref{eq:c0}). 
}
\label{Fig:Mr}
\end{figure}

Plane Wave Impulse Approximation (PWIA) corresponds to the absence of 
rescattering: 
\ba
\psi_f(\vec r)=\exp(i \vec p \cdot \vec r)\ . 
\ea
In the distorted  DWIA formalism $\exp(i \vec p \cdot \vec r)$ is substituted by 
a wave including the effects of $\bar{p}p$ rescattering. 
We choose a simple factorized distortion $D(\vec r)$
\ba
\psi_f(\vec r)=
D(\vec r)\exp(i \vec p \cdot \vec r) \ ,
\label{eq:c1}
\ea
where 
$D(\vec r)$ is calculated as a Glauber-like eikonal 
factor: 
\ba
D(x,y,z)=\exp\Big(- i b \int_z^\infty \rho(x,y,z') dz'\ \Big)
\label{eq:c2}
\ea
where 
$\hat z\ //\ \vec p$ and $b$ is a complex number, 
whose meaning is: 
\begin{itemize}
\item  Pure real $b$: elastic rescattering potential, that may be attractive 
or repulsive depending on the relative sign of $Re(b)$ and 
$p_z$. 
\item  Pure imaginary $b$: imaginary potential causing flux absorption or flux 
creation in rescattering. 

\end{itemize}
Strictly speaking the product $b\rho(\vec r)$ is not a potential. 
$V(r)$ $\equiv$ $2 b p \rho(\vec r)$ is a true optical potential appearing 
in a linearized form of the Schrodinger equation, but for simplicity we name 
"potential" the product $b \rho(\vec r)$. 

The key function is the real function $\rho(r)$, describing 
the space distribution of 
the strength and the sign of the rescattering potential ($\rho$ may be 
negative). 
We have tested three  
families of space densities $\rho(r)$ for the rescattering potential:  
\begin{enumerate}
\item   {\bf Compact rescattering densities:} they are decreasing functions 
of $r$, as for example Woods-Saxon densities. 
This class includes imaginary-dominated potentials 
that are typical of the theory of $\bar{p}p$ low-energy interactions. 

\item   {\bf Hollow rescattering densities:} they are  very small or vanishing at 
small $r$,  large in a sub-range of 0.2-2 fm, and tend to zero 
for larger $r$. 

\item   {\bf Double-layer rescattering densities:} these are the combination of 
two potentials of class 2 with opposite sign. So we may 
have a region $r_a<r<r_b$ with a repulsive potential and a region $r_c<r<r_d$ with an attractive potential, 
or we may have  two regions, characterized by an absorbing and a generating 
imaginary potential. 
\end{enumerate}
All the potentials considered here act in a range that is typical of strong 
interactions. 
Electrostatic potentials have not been considered, 
since the short-range scheme used here is not suitable 
for analyzing phase shifts that develop at distances $\gg$ 1 fm. 

Summarizing, we calculate $F(p)$ as 
\begin{align}
&F(p)\ 
= {1\over {(2\pi)^3}}   \ \int d^3 \vec r\ e^{i \vec p \cdot \vec r} D(\vec r) M_0(r),
\label{eq:c3}
\\
&M_0(r)\ \equiv\
\int d^3 \vec p\ e^{-i \vec p \cdot \vec r} F_0(p), 
\label{eq:c4}
\end{align}
with $D(\vec r)$ defined in Eq. (\ref{eq:c2}). We observe that $F(p)$ 
does not depend on the orientation of $\vec p$ because of the choice 
of constraining the $z$-axis to the direction $\vec p$ in the 
calculation of $D(\vec r)$. 

\section{Model Results}

After testing several configurations, the results are the following. 

{\bf 1) Compact potentials}.

Spherical homogeneous potentials (constant up to a 
fixed radius), Gaussian potentials and Woods-Saxon potentials 
do not give positive results. Neither real nor imaginary potentials 
have produced periodic oscillation patterns. 
This is in contrast with 
the case of the angular distributions of nuclear physics 
(see for example \cite{Bianconi:1996ur}
where evident periodic patterns are obtained via pure 
imaginary Woods-Saxon densities). However, 
in those applications the relevant variable is the $t-$channel 
momentum exchanged in elastic scattering or rescattering, 
while here it is the relative momentum of the colliding particles, that is  
equivalent to an $s-$channel momentum. 

\begin{figure}
\resizebox{0.75\textwidth}{!}{%
\includegraphics[width=14cm]{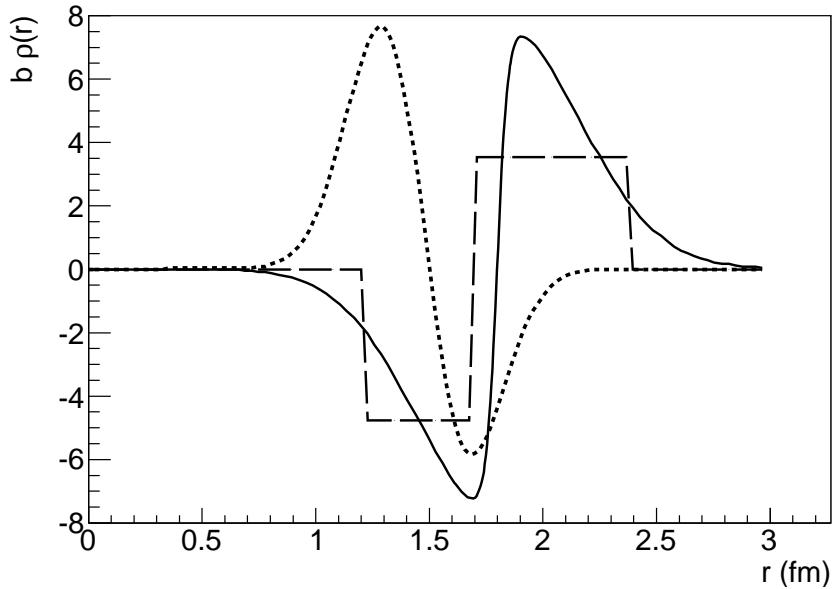}}
\caption{
The three double-layer potentials used for the fits reported 
in figures \ref{Fig:fig_fit1}, 
\ref{Fig:fig_fit2}, and \ref{Fig:fig_fit3}. 
Dashed curve: potential n.1 (Eq. (\ref{eq:potential1})). 
Continuous curve: potential 
n.2 (Eq. (\ref{eq:potential2})). 
Dotted curve: potential n.3 (Eq. (\ref{eq:potential3})). 
These potentials are used in Eq. (\ref{eq:c2}) to calculate the final state 
distortion factor $D(x,y,z)$, that leads to the fit $F(p)$ through 
Eq. (\ref{eq:c3}). 
} 
\label{Fig:potentials}
\end{figure}

{\bf 2) Hollow potentials} (not changing sign or phase in the $r$ range of interest). 

We have tested simple double-step potentials (constant between two $r$ values, 
zero elsewhere), and shifted-Gaussian potentials $~\exp[-(r-r_0)^2/\sigma^2]$, 
with both real and imaginary parts. Real hollow potentials produce periodical oscillations, but the 
oscillation period is far too large 
(2 GeV or more). In order to make it shorter,  the peak value of the potential has to be pushed to $r> 2$~ fm. 
$M_0(r)$ decreases by 3-4 orders of magnitude when 
$r$ increases by 1 fm (see Fig. \ref{Fig:Mr}).  
So, at distances $>$ 2 fm  
$M(r)$ is very small, depriving of relevance  
the effects of a real potential that is active in these regions. Imaginary potentials of pure absorbing (or pure generating) class 
have been found to be incompatible with our starting requirement that 
rescattering is a small correction. For a hollow imaginary potential 
to produce oscillations with relative magnitude 10 \%,  one needs 
a strong imaginary potential, which leading effect is damping by 
orders of magnitude the unperturbed term. 

{\bf 3) Double layer potentials} (presenting two $r$-ranges where 
the potential phase is opposite). Double-layer real potentials produce weaker oscillation effects 
than single-layer (hollow) potentials.  
Our best results have been obtained with double-layer imaginary 
potentials. These have been able to produce  
periodic oscillations with a period of 1 GeV or shorter, and of arbitrarily 
large amplitudes, depending on the parameters. Such potentials present an inner region where 
the $\bar{p}p$ flux is produced and  an outer region where  the $\bar{p}p$ 
flux is absorbed. The physical origin of this class of potentials is discussed in  next section. 

We report the results corresponding to the three different double-layer 
potentials illustrated in Fig. \ref{Fig:potentials}. All of them are purely 
imaginary ($Re(b) \equiv 0$). The former two potentials have been calibrated to 
reproduce, as well as possible, BABAR data. The third one presents 
peculiar features and non-optimal parameters, and it is reported for comparison discussion. 

{\bf Potential n.1}:  Multiple-step function
\ba
Im(b) \rho(r)=: &0\ & \mbox{ for} \ r<1\   \mbox{ fm\ and} \ r>2.4\  \mbox{ fm};\nn\\ 
&-4.8&\mbox{ for~} (1.2<r<1.7)\ \mbox{ fm};\nn\\
&3.5& \mbox{ for~} (1.7 <r<2.4)\ \mbox{ fm}; 
\label{eq:potential1}
\ea

{\bf Potential n.2}: Potential similar to the previous one, 
but regular: 
\be
Im(b) \rho(r)=B\  G(r-r_0) T(r-r_0) W(r-r_0), 
\label{eq:potential2}
\ee
where
\begin{itemize}
\item $B=7.8$ is an overall strength coefficient. 
\item $G(r-r_0)=\exp[-(r-r_0)^2/0.5^2]$ is a Gaussian with center 
in $r_0=1.8$ fm, and width $\mu=0.5$ fm. 
\item $T(r-r_0)=\tanh[(r-r_0)/0.05]$ is a "soft sign function", that is equal to $-1$ for $r\ll r_0$ fm, to $+1$ for $r\gg r_0$ fm, and changes smoothly sign in a range of 0.1 fm. 
\item $W(r-r_0)=1 + 0.05 (r-r_0)$ 
is a weight function that (slightly) increases the strength of the 
external absorbing peak over the internal generating one. 
\end{itemize}
{\bf Potential n.3}: Sum of a positive Gaussian in the inner region  and a 
negative one in the outer region: 
\be
Im(b) \rho(r)=A^+ G^+(r-r^+) - A^- G^-(r-r^-),
\label{eq:potential3}
\ee
with   
$G^\pm \equiv \exp(-[(r-r^\pm)^2/w^2]$, $r^+=$ 1.35 fm, 
$r^-= r^++w$, and 
$w=$ 0.26 fm, $A^+$=10, $A^-$= 8.4. 

This potential is very different from the previous two: the sign of the 
inner and outer parts are reversed (absorption inside), the average 
radius is smaller, the negative and positive peaks are more distant. 
It is reported here  to highlight some effects of the parameters 
rather than for good fit purposes.

\begin{figure}
\resizebox{0.75\textwidth}{!}{%
\includegraphics[width=14cm]{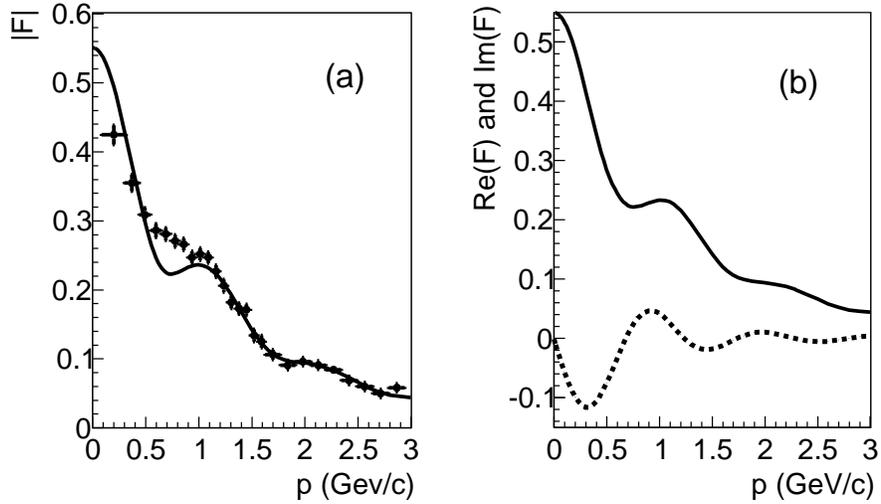}}
\caption{(a): Continuous curve: $|F(p)|$, 
obtained with the double-layer rescattering potential n.1 (the  
multiple-step function in Fig. \ref{Fig:potentials}, see text), 
compared  to the BABAR data points (full circles). 
(b):  real (solid line)  and imaginary (dashed line) parts of 
the model $F(p)$.
} 
\label{Fig:fig_fit1}
\end{figure}

\begin{figure}
\resizebox{0.75\textwidth}{!}{%
\includegraphics[width=14cm]{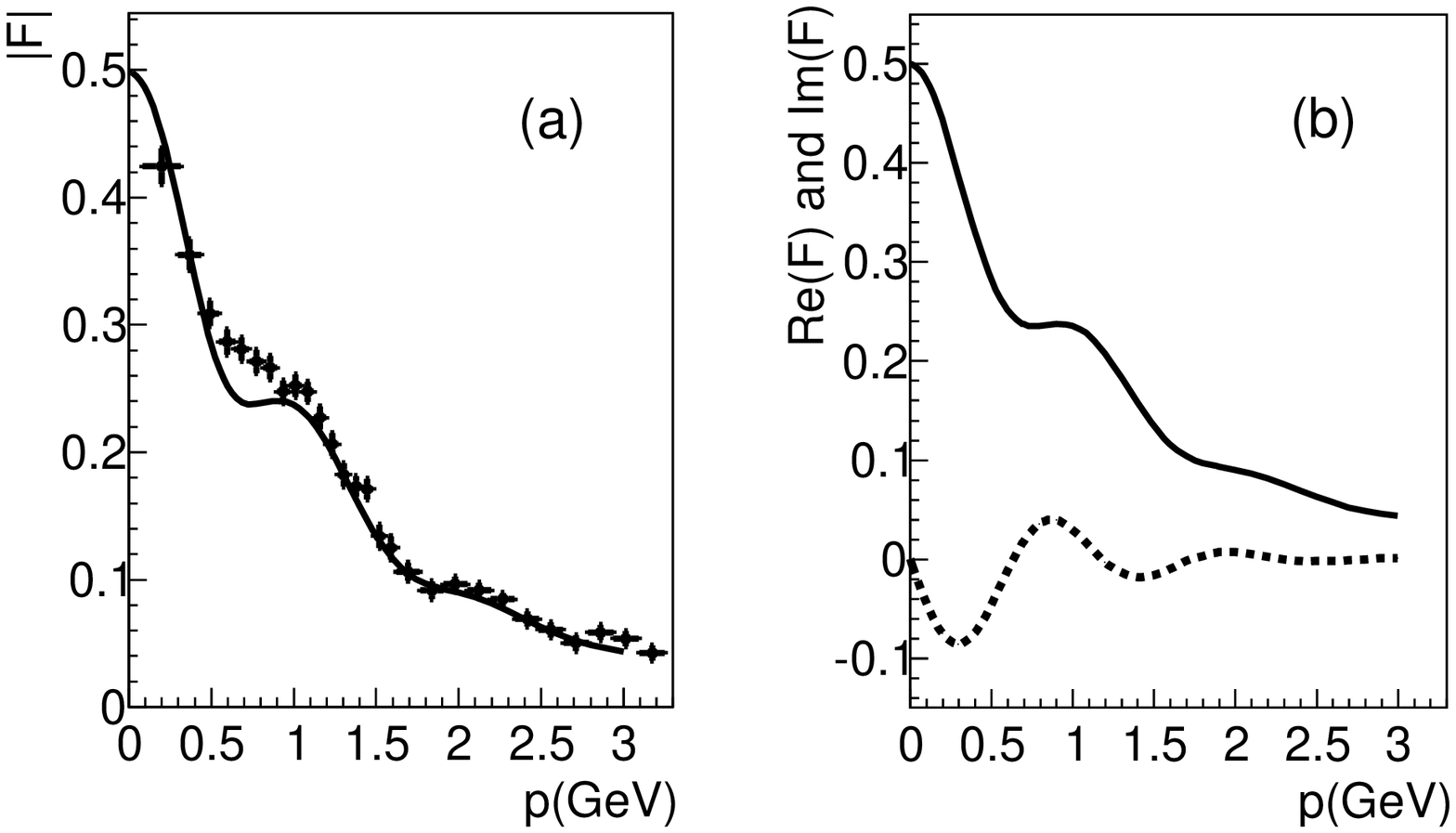}}
\caption{(a): Continuous curve: $|F(p)|$, 
obtained with the double-layer rescattering potential n.2 (continuous curve 
in Fig. \ref{Fig:potentials}, see text), 
compared  to the BABAR data points (full circles). 
(b): real (solid line)  and imaginary (dashed line) parts of 
the model $F(p)$.
} 
\label{Fig:fig_fit2}
\end{figure}

\begin{figure}
\resizebox{0.75\textwidth}{!}{%
\includegraphics[width=14cm]{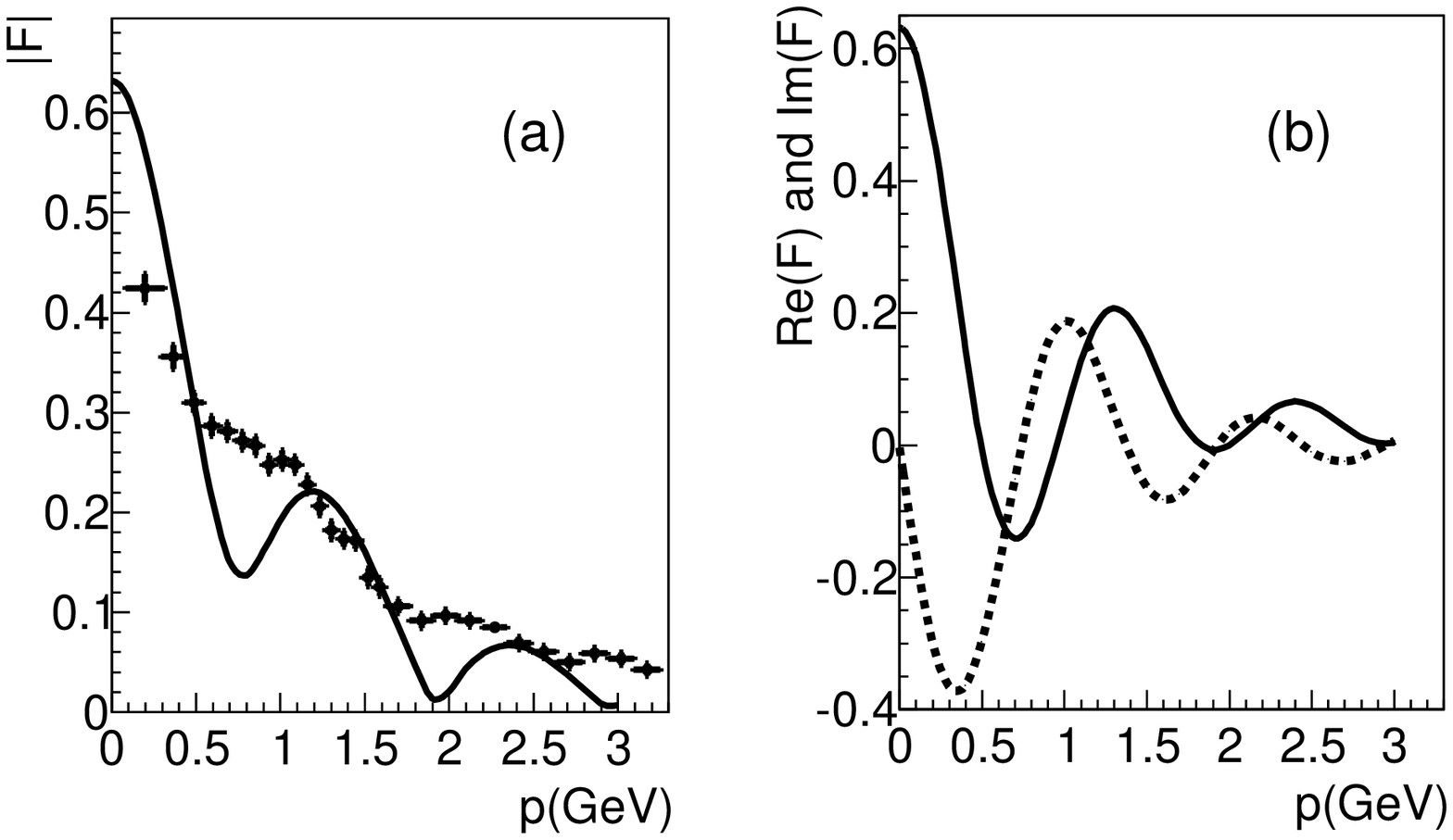}}
\caption{(a):  Continuous curve: $|F(p)|$, 
obtained with the double-layer rescattering potential n.3 (continuous 
dotted curve in Fig. \ref{Fig:potentials}, see text), compared  
to the BABAR data points (full circles). 
(b):  real (solid line)  and imaginary (dashed line) parts of $F(p)$.
} 
\label{Fig:fig_fit3}
\end{figure}

The results obtained with these potentials are presented in Figs. \ref{Fig:fig_fit1}, 
\ref{Fig:fig_fit2}, and \ref{Fig:fig_fit3}. 
The fits with potentials 1 and 2 reproduce satisfactorily well the data, given the simplicity of the model. In these two cases, the data are slightly overestimated near 
the threshold. This 
may be attributed to the small-energy limitations of the 
eikonal formalism chosen here 
to reproduce the wave distortion. As observed 
in \cite{Bianconi:1996ur} the use of 
this approximation within DWIA is good when several partial waves 
are involved in rescattering, and definitely it does not apply in a regime of 
S-wave dominance, corresponding to $p\le 200$~MeV  for $\bar{p}p$ systems \cite{Bianconi:2000ap}. 

The example with potential n.3 shows that 
it is possible to obtain similar qualitative results 
with opposite configurations: absorbing 
potential  in the outer region and generating potential in the inner region, or viceversa. 
For our best fits we have preferred the first option, because it corresponds to the phenomenology of $\bar{p}p$ annihilation, dominated by flux absorption when the proton 
and antiproton begin to overlap. 

Potential n.3 allows for an easy analysis of the separate role 
of the two potential layers, since one may independently 
modify the peak strengths $A^+$ and $A^-$. Systematic attempts show that it 
is possible to obtain periodic oscillations of pretty large amplitude by 
increasing both $A^+$ and $A^-$, at the condition that 
the relative effect of the absorbing and of the creating 
parts of the potential is well balanced, that may be obtained by acting on the ratio $A^+/A^-$. 
Apart for avoiding normalization problems, an equilibrated balance between the two strengths 
is one of the keys to get remarkable and periodic oscillations. 
 
On the other side, this potential is not suitable for producing 
arbitrarily short oscillation periods, because it lacks a decisive 
feature of potentials n.1 and 2: 
their steep derivative at the point where they change sign. To obtain 
this feature with potential n.3, the distance between the two peaks must be smaller than 
their width. In such conditions the two Gaussians overlap 
and cancel reciprocally. We have been able to reduce the oscillation 
period down to what is visible in Fig. \ref{Fig:fig_fit3}, but not further. 
The conclusion is that the period of the oscillations is related to a sudden transition between the 
flux feeding and the flux depleting regions. 

Another important property shared by the  three double-layer potentials is to 
produce a systematic threshold enhancement:  $p=0$  corresponds to an oscillation 
maximum, if the effect of the flux-creating and flux-absorbing parts of 
the potential are reasonably well balanced.
This property is very stable, it is not related 
to a special set of parameters, and does not depend on the fact that the 
absorbing part of the potential is external (potentials n.1 and n.2) 
or internal (potential n.3). So, the threshold enhancement is an intrinsic  
property of the imaginary double-layer model.

For large $r$, potentials n.1 and n.2 act qualitatively 
as the purely absorptive potentials used to fit $\bar{p}p$ LEAR 
data \cite{Bruckner:1989ew,Bianconi:2000ap,Batty:2000vr,Friedman:2014vva}.  
To reproduce LEAR elastic and annihilation data, 
phenomena taking place at small $r$ have no relevance, since  
the surface interaction at $r \approx$ 1-2 fm 
prevents most of the initial $\bar{p}p$ channel wave function 
from entering the $r<1$~ fm region. The "regeneration'' effect  
due to the inner potential introduced above would have little  effect on 
the total elastic and inelastic cross sections, since it
affects only a very small component of the wave function.
 
On the other side, this small component  which is nonzero  
at  small $r$ is essential for the coupling 
of the proton-antiproton pair with a virtual photon with virtuality $q^2>4M^2$. The coupled regeneration/absorption mechanism introduced here  
produces, for $r<2$ fm, 
an alternance of regions where this component of the 
$\bar{p}p$ wave function is enhanced or suppressed. Let us discuss the 
conditions leading to observable effects. 
The  enhancement of the cross sections at small $p$, where the 
the $\bar{p}p$ distorted wave function 
$\exp[i \vec p \cdot \vec r]D(\vec r)$ approximately reduces to 
$D(\vec r)$ and the Fourier transform 
of Eq. (\ref{eq:c3}) simplifies to $\int d^3r D(\vec r) M_0(r)$,   
suggests that the potential enhances the wave function 
at small $r$ where $M_0(r)$ is very large (see Fig. \ref{Fig:Mr}). 
This effect is not specific of double-layer potentials: 
for example with a spherical real attracting potential does the same. 
The presence of further 
oscillations at larger $p$ however suggests that  
double-layer imaginary potentials create regions 
where the product $D(\vec r) M_0(r)$ alternatively 
becomes larger and smaller, enough to "resonate'' with the 
Fourier transform factor $\exp(i \vec p \cdot \vec r)$ for periodic 
$p$ values far from the threshold (see Fig. 3b of \cite{Bianconi:2015owa}). 
If two regions with a larger 
and a smaller value of $D(\vec r)$ are present at $r_+$ and $r_-$ 
respectively, $r_+-r_-$ must be small, 
or the steep $r$-decrease of $M_0(r)$ will 
make the modulation by $D(r_+)$ negligible with respect to $D(r_-)$. 
This may explain the 
relevance of having a steep potential at the change of sign. 

\section{Hypothesis on the physical origin of  
"inner creative/ outer absorptive" potentials}

An optical potential with an imaginary part may be justified within 
several theoretical frameworks but in general, and intuitively, its origin 
is related to the fact that a a multi-channel 
process is inclusively projected onto one channel alone. 


\begin{figure}
\mbox{\epsfxsize=10.cm\leavevmode \epsffile{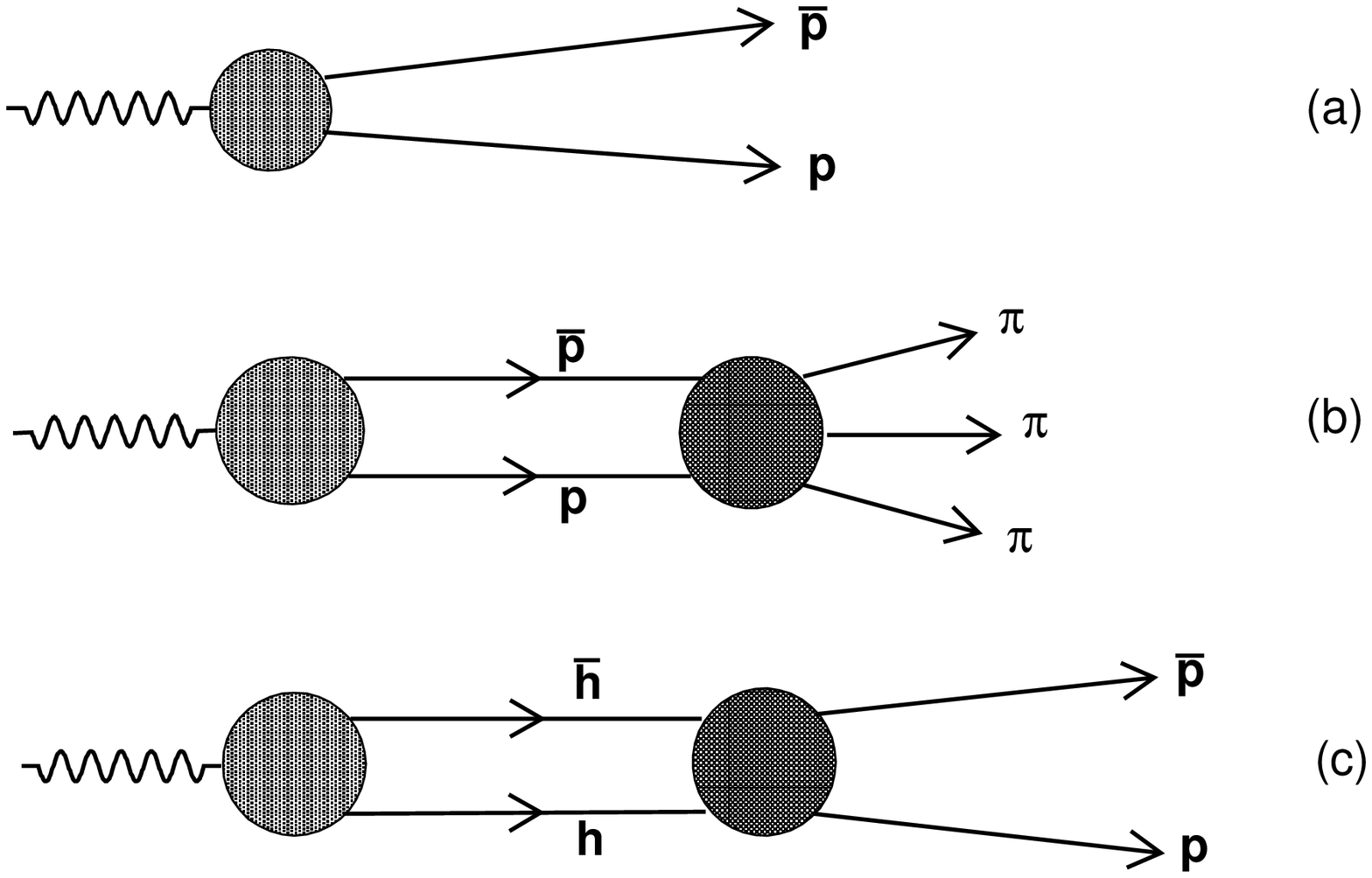}}
\caption{Examples of diagrams entering the absorptive and creative parts 
of the potential. 
(a): ``direct'' $\gamma^*$ $\rightarrow$ $\bar{p}p$. 
The light-grey circle represents a model 
amplitude without contributions by rescattering. It 
leads to the background regular term $F_0(p)$ or 
equivalently, 
in $r-$coordinate representation, to $M_0(r)$ (see Eq. (\ref{eq:c0})). 
(b): The $\gamma^*$ $\rightarrow$ $\bar{p}p$ production  
is followed by an annihilation 
process reducing the final $\bar{p}p$ outcome. This contributes to 
the absorptive part of the optical potential. 
(c): The same model previously used to calculate the direct production 
$\gamma^*$ $\rightarrow$ $\bar{p}p$ is now used to calculate 
$\gamma^*$ $\rightarrow$ $\bar{h}h$, where $h$ is a hadron different from 
a proton. Rescattering converts $\bar{h}h$ into 
$\bar{p}p$ increasing the final output. The effect of this diagram is 
taken into account by the creative part of the potential. Even  
other diagrams, with intermediate hadronic states more complicated than 
$\bar{h}h$, may contribute.  
}
\label{Fig:rescattering}
\end{figure}


For the case of interest, this  
is illustrated in Fig. \ref{Fig:rescattering}. Diagram (a) 
is the amplitude of $\gamma^*$ $\rightarrow$ $\bar{p}p$ 
within a model that does not include $\bar{p}p$ rescattering.  
This is supposed to lead to the background regular 
component of the form factor, without oscillations. Diagram (b) 
considers the possibility that $\bar{p}p$ annihilation into a multi-meson 
state depletes the final state produced by process (a). In our formalism 
this finds an expression in the absorption component of the 
imaginary potential. The 
most interesting additional diagram is (c). 
The same model that in case (a) has been used to calculate the 
amplitude of $\gamma^*$ $\rightarrow$ $\bar{p}p$, is used in case 
(c) to calculate the amplitude of $\gamma^*$ $\rightarrow$ $\bar{h}h$, 
where $h$ is a hadron that is different from $p$, for example a neutron 
or a higher mass baryon. Later rescattering converts this pair into 
a $\bar{p}p$ state. The intermediate state 
is not necessarily a two-particle state. Any multi-meson 
state with the right quantum numbers 
may play the role of an intermediate state that is later converted 
into $\bar{p}p$. 

According with the previous argument, the double-layer optical potential 
used here is not in conflict with the existing models for the TL FF, 
but is rather 
an effective way to include rescattering corrections to these models. 
Many models for the hadron coupling to the virtual photon have been 
developed and applied to the calculation of SL FFs. 
Some of them may be analytically 
continued to the TL region. This is the case for approaches based on 
vector meson dominance  \cite{Bijker:2004yu, Adamuscin:2005aq} and 
dispersion relations  \cite{Belushkin:2006qa,Lomon:2012pn}.  
Constituent quark models in light front dynamics 
may be applied  \cite{deMelo:2003uk}, as well as approaches based on 
AdS/QCD correspondence \cite{Brodsky:2007hb}. A phenomenological picture 
for the full time-evolution of the hadronization process
has been proposed in \cite{Kuraev:2011vq}.

Practically all these models may be the starting point for a calculation 
within the proposed DWIA-optical 
potential scheme, following the prescription suggested in 
fig.\ref{Fig:rescattering}: 

Step 1) The model is used to calculate  
the PWIA production amplitude of the $\bar{p}p$ pair (diagram (a)), 
and this leads the background regular component of the 
effective form factor, and to its Fourier transform $M_0(r)$ in 
coordinate representation (Eq. (\ref{eq:c0})). 

Step 2) Final state processes implying he annihilation of the $\bar{p}p$ pair 
into mesons are added (diagram (b)). 
The relevant amplitudes of this group 
may be effectively summarized in a flux-absorbing 
optical potential that in coordinate representation modifies  
the plane wave of the $\bar{p}p$ channel as in 
Eqs. (\ref{eq:c1}) and (\ref{eq:c2}). 

Step 3) The model is used for calculating the amplitude for the 
production of other hadronic states $\bar{h}h$ 
that are later converted into $\bar{p}p$ by rescattering (diagram (c)). 
The amplitudes for the processes of this group may be 
effectively summarized in a flux-creating optical potential 
distorting the plane wave of the $\bar{p}p$ channel. 

In principle processes as in Figs. \ref{Fig:rescattering}b and \ref{Fig:rescattering}c are 
possible everywhere in a range of a few fm around the initial virtual photon 
decay point. Why should the ``flux enhancing'' diagrams like (c) dominate 
the small$-r$ regions? 

While the explanation of the optical potential in terms of multi-step 
inelastic reactions is straightforward, 
for the answer to this question we may 
only propose an hypothesis, that relates the presence of the 
creation part of the potential to those regions where high-mass 
virtual intermediate states are more likely.  

The amplitude for the transition from $\bar{p}p$ to a state made 
of 3-10 mesons is not different from the amplitude for the reverse 
process, but phase space 
makes the probability of the former process larger than the probability 
of the latter. So, 
the hadronic states that may contribute to feeding the $\bar{p}p$ channel 
(Fig. \ref{Fig:rescattering}c) and 
not to further depleting it (Fig. \ref{Fig:rescattering}b), are the states made by one or 
two heavy hadrons like $N^*\bar{N}^*$ 
states. Unless $q$ $>>$ $2M_p$, the hadrons composing these states 
are virtual, short-lived and slow, with few exceptions 
like a neutron-antineutron intermediate state. 
So, they play a role 
for small $r$ only, since small $r$ corresponds in the average 
to small times after the 
photon conversion into the first $\bar{q}q$ pair. 

On the other side, these high-mass states must be present in the state 
that is initially produced by the decay of a virtual photon with 
$q$ $\geq$ $2M_p$ according to the statement that this 
state has space-time size of magnitude $1/q$. 
According to the PQCD view\cite{Brodsky:1973kr}, in the SL case  
(elastic electron-proton scattering) the virtual photon is absorbed by a 
fluctuation of the proton state consisting of 
valence quarks grouped within a space-time region of size $1/q$. 
The fact that this fluctuation exists means that in the TL case 
a corresponding fluctuation of the $\bar{p}p$ state exists where 
the required number of valence quarks and antiquarks 
is concentrated within a region of space-time size $1/q$. Indeed, the 
Feynman diagrams describing the PQCD kernel of the process 
are the same in SL and TL and in these diagrams  
all the involved partons are connected by propagator 
lines with off-shellness of magnitude $q$, that obliges them to be 
within a space-time distance $1/q$. 

If this $1/q$-sized fluctuation takes place in a $\bar{p}p$ annihilation, 
we may have the rare but possible event 
$\bar{p}p$ $\rightarrow$ $e^+e^-$. 
In the reaction 
$e^+e^-$ $\rightarrow$ $\bar{p}p$, the path is opposite:  
the virtual 
photon creates a $1/q$-sized fluctuation of quarks and antiquarks, 
that may evolve into a $\bar{p}p$ pair, 
but may also evolve into other hadronic states (e.g. neutron-antineutron) 
since also these states present $1/q$-sized fluctuations of their 
parton content. 

Any configuration of a color singlet state, like the 
``3 quark + 3 antiquark'' small-sized state produced by the decay of the 
virtual photon, may be written as a sum over physical hadronic states 
with the same quantum numbers, since these states form a complete basis 
for this system. However, a state with a size 
of magnitude $1/q$ cannot be reproduced by the sum of a small number of 
hadronic states since these have a typical size 1 fm. What is needed 
is a set of several states which interfere destructively at distances 
$>$ $1/q$ from the virtual photon materialization point, and 
constructively at distances $\lesssim$ $1/q$, so to build a wave packet 
of size $1/q$. 
Taking into account that $1/q$ is also the magnitude of the 
lifetime of this fluctuation, we may estimate that 
the sum must include hadronic states with a spread of magnitude $q$ 
in their center of mass energy. 
With a virtuality that can be of the 
same magnitude as $q$, it is evident that many of these states cannot 
propagate far from the virtual photon materialization point, and 
this may support the dominance of the flux-enhancing term of the optical 
potential at small $r$. 

As observed, this picture behind the small-$r$ dominance of the 
flux-creation part of the potential is just an educated guess, 
because of the difficulties in passing from qualitative ideas to a 
detailed model.

\section{Conclusions}

We have analyzed the modulation structure shown by the precise data on 
the TL proton form factor, recently obtained by the BABAR 
collaboration. First, we have repeated 
the data analysis already presented in our previous 
work \cite{Bianconi:2015owa} for the case of four different  
form factor parameterizations available in the literature. 
The difference between BABAR data and the form factor 
parametrization is well fitted by an oscillating function of the form 
$A~ \exp(-Bp) ~\cos(Cp)$, where $p$ is the momentum of the relative motion 
of the $\bar{p}p$ pair. The periodicity of 
the $\cos(Cp)$ term is verified within 15 \% in a $p$ range from zero to 
2.8 GeV. 

The periodicity of this oscillating modulation as a function of the 
relative momentum of the final hadrons 
has been qualitatively explained in terms of rescattering between the 
final products of the reaction 
$e^+e^-$ $\rightarrow$ $\bar{p}p$ and 
reproduced via an optical potential of peculiar (double spherical layer) 
form. 

An imaginary optical potential that is mainly 
flux-generating in a region of small distances between the centers 
of the forming (and still overlapping) proton and antiproton, and mainly 
flux-absorbing at larger distances, produces systematic oscillations 
of the effective proton TL form factor, consistent with  the observed ones. At 
distances $\approx$ 1-2 fm such a potential 
behaves as the optical potentials ordinarily used to reproduce 
$\bar{p}p$ annihilation data, that is it damps the $\bar{p}p$ flux by 
annihilating $\bar{p}$ and $p$ into multi-meson states. 
A possible explanation for the regeneration features of the potential at smaller distances could be 
in terms of coupling between the  $\bar{p}p$ final channel and large-mass virtual states (like 
baryon-antibaryon) temporarily produced by the virtual photon. 
In order to reproduce the data, the transition from the flux-generating
to the flux-absorbing region must be sudden. 
A soft transition produces oscillations with periods longer 
than the observed one. With this double-layer structure, we always find 
threshold enhancement of the form factors. So, within this scheme threshold enhancement 
and oscillations are expressions of the same phenomenon. 

We have tested other simpler  configurations  of the potential, and also real potentials with 
a range typical of strong interactions, but these do not seem to allow 
for oscillations with the required period and strength. 
The proposed phenomenological scheme is compatible with existing 
theoretical models for the TL form factors, since it may be considered
as a rescattering correction that does not touch the core schemes 
of these models.


\begin{thebibliography}{39}
\expandafter\ifx\csname natexlab\endcsname\relax\def\natexlab#1{#1}\fi
\expandafter\ifx\csname bibnamefont\endcsname\relax
  \def\bibnamefont#1{#1}\fi
\expandafter\ifx\csname bibfnamefont\endcsname\relax
  \def\bibfnamefont#1{#1}\fi
\expandafter\ifx\csname citenamefont\endcsname\relax
  \def\citenamefont#1{#1}\fi
\expandafter\ifx\csname url\endcsname\relax
  \def\url#1{\texttt{#1}}\fi
\expandafter\ifx\csname urlprefix\endcsname\relax\def\urlprefix{URL }\fi
\providecommand{\bibinfo}[2]{#2}
\providecommand{\eprint}[2][]{\url{#2}}

\bibitem[{\citenamefont{Pacetti et~al.}(2015)\citenamefont{Pacetti,
  Baldini~Ferroli, and Tomasi-Gustafsson}}]{Pacetti:2015iqa}
\bibinfo{author}{\bibfnamefont{S.}~\bibnamefont{Pacetti}},
  \bibinfo{author}{\bibfnamefont{R.}~\bibnamefont{Baldini~Ferroli}},
  \bibnamefont{and}
  \bibinfo{author}{\bibfnamefont{E.}~\bibnamefont{Tomasi-Gustafsson}},
  \bibinfo{journal}{Phys.Rep.} \textbf{\bibinfo{volume}{550-551}},
  \bibinfo{pages}{1} (\bibinfo{year}{2015}).

\bibitem[{\citenamefont{Zichichi et~al.}(1962)\citenamefont{Zichichi, Berman,
  Cabibbo, and Gatto}}]{Zichichi:1962ni}
\bibinfo{author}{\bibfnamefont{A.}~\bibnamefont{Zichichi}},
  \bibinfo{author}{\bibfnamefont{S.}~\bibnamefont{Berman}},
  \bibinfo{author}{\bibfnamefont{N.}~\bibnamefont{Cabibbo}}, \bibnamefont{and}
  \bibinfo{author}{\bibfnamefont{R.}~\bibnamefont{Gatto}},
  \bibinfo{journal}{Nuovo Cim.} \textbf{\bibinfo{volume}{24}},
  \bibinfo{pages}{170} (\bibinfo{year}{1962}).

\bibitem[{\citenamefont{Bardin et~al.}(1994)\citenamefont{Bardin, Burgun,
  Calabrese, Capon, Carlin et~al.}}]{Bardin:1994am}
\bibinfo{author}{\bibfnamefont{G.}~\bibnamefont{Bardin}},
  \bibinfo{author}{\bibfnamefont{G.}~\bibnamefont{Burgun}},
  \bibinfo{author}{\bibfnamefont{R.}~\bibnamefont{Calabrese}},
  \bibinfo{author}{\bibfnamefont{G.}~\bibnamefont{Capon}},
  \bibinfo{author}{\bibfnamefont{R.}~\bibnamefont{Carlin}},
  \bibnamefont{et~al.}, \bibinfo{journal}{Nucl.Phys.}
  \textbf{\bibinfo{volume}{B411}}, \bibinfo{pages}{3} (\bibinfo{year}{1994}).

\bibitem[{\citenamefont{Balestra et~al.}(1986)}]{Balestra:1985kn}
\bibinfo{author}{\bibfnamefont{F.}~\bibnamefont{Balestra}}
  \bibnamefont{et~al.}, \bibinfo{journal}{Nucl. Phys.}
  \textbf{\bibinfo{volume}{A452}}, \bibinfo{pages}{573} (\bibinfo{year}{1986}).

\bibitem[{\citenamefont{Balestra et~al.}(1989)}]{Balestra:1989rb}
\bibinfo{author}{\bibfnamefont{F.}~\bibnamefont{Balestra}}
  \bibnamefont{et~al.}, \bibinfo{journal}{Phys. Lett.}
  \textbf{\bibinfo{volume}{B230}}, \bibinfo{pages}{36} (\bibinfo{year}{1989}).

\bibitem[{\citenamefont{Bizzarri et~al.}(1974)\citenamefont{Bizzarri, Guidoni,
  Marcelja, Marzano, Castelli, and Sessa}}]{Bizzarri:1973sp}
\bibinfo{author}{\bibfnamefont{R.}~\bibnamefont{Bizzarri}},
  \bibinfo{author}{\bibfnamefont{P.}~\bibnamefont{Guidoni}},
  \bibinfo{author}{\bibfnamefont{F.}~\bibnamefont{Marcelja}},
  \bibinfo{author}{\bibfnamefont{F.}~\bibnamefont{Marzano}},
  \bibinfo{author}{\bibfnamefont{E.}~\bibnamefont{Castelli}}, \bibnamefont{and}
  \bibinfo{author}{\bibfnamefont{M.}~\bibnamefont{Sessa}},
  \bibinfo{journal}{Nuovo Cim.} \textbf{\bibinfo{volume}{A22}},
  \bibinfo{pages}{225} (\bibinfo{year}{1974}).

\bibitem[{\citenamefont{Bruckner et~al.}(1990)\citenamefont{Bruckner, Cujec,
  Dobbeling, Dworschak, Guttner et~al.}}]{Bruckner:1989ew}
\bibinfo{author}{\bibfnamefont{W.}~\bibnamefont{Bruckner}},
  \bibinfo{author}{\bibfnamefont{B.}~\bibnamefont{Cujec}},
  \bibinfo{author}{\bibfnamefont{H.}~\bibnamefont{Dobbeling}},
  \bibinfo{author}{\bibfnamefont{K.}~\bibnamefont{Dworschak}},
  \bibinfo{author}{\bibfnamefont{F.}~\bibnamefont{Guttner}},
  \bibnamefont{et~al.}, \bibinfo{journal}{Z.Phys.}
  \textbf{\bibinfo{volume}{A335}}, \bibinfo{pages}{217} (\bibinfo{year}{1990}).

\bibitem[{\citenamefont{Balestra et~al.}(1984)}]{Balestra:1984wd}
\bibinfo{author}{\bibfnamefont{F.}~\bibnamefont{Balestra}}
  \bibnamefont{et~al.}, \bibinfo{journal}{Phys. Lett.}
  \textbf{\bibinfo{volume}{B149}}, \bibinfo{pages}{69} (\bibinfo{year}{1984}).

\bibitem[{\citenamefont{Balestra et~al.}(1985)}]{Balestra:1985wk}
\bibinfo{author}{\bibfnamefont{F.}~\bibnamefont{Balestra}}
  \bibnamefont{et~al.}, \bibinfo{journal}{Phys. Lett.}
  \textbf{\bibinfo{volume}{B165}}, \bibinfo{pages}{265} (\bibinfo{year}{1985}).

\bibitem[{\citenamefont{Bertin et~al.}(1996)}]{Bertin:1996kw}
\bibinfo{author}{\bibfnamefont{A.}~\bibnamefont{Bertin}} \bibnamefont{et~al.}
  (\bibinfo{collaboration}{OBELIX Collaboration}), \bibinfo{journal}{Phys.
  Lett.} \textbf{\bibinfo{volume}{B369}}, \bibinfo{pages}{77}
  (\bibinfo{year}{1996}).

\bibitem[{\citenamefont{Benedettini et~al.}(1997)}]{Benedettini:1997fk}
\bibinfo{author}{\bibfnamefont{A.}~\bibnamefont{Benedettini}}
  \bibnamefont{et~al.} (\bibinfo{collaboration}{OBELIX Collaboration}),
  \bibinfo{journal}{Nucl. Phys. Proc. Suppl.} \textbf{\bibinfo{volume}{56}},
  \bibinfo{pages}{58} (\bibinfo{year}{1997}).

\bibitem[{\citenamefont{Zenoni et~al.}(1999{\natexlab{a}})\citenamefont{Zenoni,
  Bianconi, Bonomi, Corradini, Donzella et~al.}}]{Zenoni:1999su}
\bibinfo{author}{\bibfnamefont{A.}~\bibnamefont{Zenoni}},
  \bibinfo{author}{\bibfnamefont{A.}~\bibnamefont{Bianconi}},
  \bibinfo{author}{\bibfnamefont{G.}~\bibnamefont{Bonomi}},
  \bibinfo{author}{\bibfnamefont{M.}~\bibnamefont{Corradini}},
  \bibinfo{author}{\bibfnamefont{A.}~\bibnamefont{Donzella}},
  \bibnamefont{et~al.}, \bibinfo{journal}{Phys.Lett.}
  \textbf{\bibinfo{volume}{B461}}, \bibinfo{pages}{413}
  (\bibinfo{year}{1999}{\natexlab{a}}).

\bibitem[{\citenamefont{Zenoni et~al.}(1999{\natexlab{b}})\citenamefont{Zenoni,
  Bianconi, Bocci, Bonomi, Corradini et~al.}}]{Zenoni:1999st}
\bibinfo{author}{\bibfnamefont{A.}~\bibnamefont{Zenoni}},
  \bibinfo{author}{\bibfnamefont{A.}~\bibnamefont{Bianconi}},
  \bibinfo{author}{\bibfnamefont{F.}~\bibnamefont{Bocci}},
  \bibinfo{author}{\bibfnamefont{G.}~\bibnamefont{Bonomi}},
  \bibinfo{author}{\bibfnamefont{M.}~\bibnamefont{Corradini}},
  \bibnamefont{et~al.}, \bibinfo{journal}{Phys.Lett.}
  \textbf{\bibinfo{volume}{B461}}, \bibinfo{pages}{405}
  (\bibinfo{year}{1999}{\natexlab{b}}).

\bibitem[{\citenamefont{Bianconi
  et~al.}(2000{\natexlab{a}})\citenamefont{Bianconi, Bonomi, Bussa,
  Lodi~Rizzini, Venturelli et~al.}}]{Bianconi:2000nh}
\bibinfo{author}{\bibfnamefont{A.}~\bibnamefont{Bianconi}},
  \bibinfo{author}{\bibfnamefont{G.}~\bibnamefont{Bonomi}},
  \bibinfo{author}{\bibfnamefont{M.}~\bibnamefont{Bussa}},
  \bibinfo{author}{\bibfnamefont{E.}~\bibnamefont{Lodi~Rizzini}},
  \bibinfo{author}{\bibfnamefont{L.}~\bibnamefont{Venturelli}},
  \bibnamefont{et~al.}, \bibinfo{journal}{Phys.Lett.}
  \textbf{\bibinfo{volume}{B481}}, \bibinfo{pages}{194}
  (\bibinfo{year}{2000}{\natexlab{a}}).

\bibitem[{\citenamefont{Bianconi
  et~al.}(2000{\natexlab{b}})\citenamefont{Bianconi, Bonomi, Lodi~Rizzini,
  Venturelli, and Zenoni}}]{Bianconi:1999vq}
\bibinfo{author}{\bibfnamefont{A.}~\bibnamefont{Bianconi}},
  \bibinfo{author}{\bibfnamefont{G.}~\bibnamefont{Bonomi}},
  \bibinfo{author}{\bibfnamefont{E.}~\bibnamefont{Lodi~Rizzini}},
  \bibinfo{author}{\bibfnamefont{L.}~\bibnamefont{Venturelli}},
  \bibnamefont{and} \bibinfo{author}{\bibfnamefont{A.}~\bibnamefont{Zenoni}},
  \bibinfo{journal}{Phys.Rev.} \textbf{\bibinfo{volume}{C62}},
  \bibinfo{pages}{014611} (\bibinfo{year}{2000}{\natexlab{b}}).

\bibitem[{\citenamefont{Bianconi et~al.}(2011)\citenamefont{Bianconi,
  Corradini, Hori, Leali, Lodi~Rizzini et~al.}}]{Bianconi:2011zz}
\bibinfo{author}{\bibfnamefont{A.}~\bibnamefont{Bianconi}},
  \bibinfo{author}{\bibfnamefont{M.}~\bibnamefont{Corradini}},
  \bibinfo{author}{\bibfnamefont{M.}~\bibnamefont{Hori}},
  \bibinfo{author}{\bibfnamefont{M.}~\bibnamefont{Leali}},
  \bibinfo{author}{\bibfnamefont{E.}~\bibnamefont{Lodi~Rizzini}},
  \bibnamefont{et~al.}, \bibinfo{journal}{Phys.Lett.}
  \textbf{\bibinfo{volume}{B704}}, \bibinfo{pages}{461} (\bibinfo{year}{2011}).

\bibitem[{\citenamefont{Bianconi
  et~al.}(2000{\natexlab{c}})\citenamefont{Bianconi, Bonomi, Bussa,
  Lodi~Rizzini, Venturelli et~al.}}]{Bianconi:2000ap}
\bibinfo{author}{\bibfnamefont{A.}~\bibnamefont{Bianconi}},
  \bibinfo{author}{\bibfnamefont{G.}~\bibnamefont{Bonomi}},
  \bibinfo{author}{\bibfnamefont{M.}~\bibnamefont{Bussa}},
  \bibinfo{author}{\bibfnamefont{E.}~\bibnamefont{Lodi~Rizzini}},
  \bibinfo{author}{\bibfnamefont{L.}~\bibnamefont{Venturelli}},
  \bibnamefont{et~al.}, \bibinfo{journal}{Phys.Lett.}
  \textbf{\bibinfo{volume}{B483}}, \bibinfo{pages}{353}
  (\bibinfo{year}{2000}{\natexlab{c}}).

\bibitem[{\citenamefont{Batty et~al.}(2001)\citenamefont{Batty, Friedman, and
  Gal}}]{Batty:2000vr}
\bibinfo{author}{\bibfnamefont{C.}~\bibnamefont{Batty}},
  \bibinfo{author}{\bibfnamefont{E.}~\bibnamefont{Friedman}}, \bibnamefont{and}
  \bibinfo{author}{\bibfnamefont{A.}~\bibnamefont{Gal}},
  \bibinfo{journal}{Nucl.Phys.} \textbf{\bibinfo{volume}{A689}},
  \bibinfo{pages}{721} (\bibinfo{year}{2001}).

\bibitem[{\citenamefont{Friedman}(2014)}]{Friedman:2014vva}
\bibinfo{author}{\bibfnamefont{E.}~\bibnamefont{Friedman}},
  \bibinfo{journal}{Nucl.Phys.} \textbf{\bibinfo{volume}{A925}},
  \bibinfo{pages}{141} (\bibinfo{year}{2014}).

\bibitem[{\citenamefont{Lees et~al.}(2013{\natexlab{a}})}]{Lees:2013xe}
\bibinfo{author}{\bibfnamefont{J.}~\bibnamefont{Lees}} \bibnamefont{et~al.}
  (\bibinfo{collaboration}{BaBar Collaboration}), \bibinfo{journal}{Phys.Rev.}
  \textbf{\bibinfo{volume}{D87}}, \bibinfo{pages}{092005}
  (\bibinfo{year}{2013}{\natexlab{a}}).

\bibitem[{\citenamefont{Lees et~al.}(2013{\natexlab{b}})}]{Lees:2013uta}
\bibinfo{author}{\bibfnamefont{J.}~\bibnamefont{Lees}} \bibnamefont{et~al.}
  (\bibinfo{collaboration}{BaBar Collaboration}), \bibinfo{journal}{Phys.Rev.}
  \textbf{\bibinfo{volume}{D88}}, \bibinfo{pages}{072009}
  (\bibinfo{year}{2013}{\natexlab{b}}).

\bibitem[{\citenamefont{Bianconi and
  Tomasi-Gustafsson}(2015)}]{Bianconi:2015owa}
\bibinfo{author}{\bibfnamefont{A.}~\bibnamefont{Bianconi}} \bibnamefont{and}
  \bibinfo{author}{\bibfnamefont{E.}~\bibnamefont{Tomasi-Gustafsson}},
  \bibinfo{journal}{Phys. Rev. Lett.} \textbf{\bibinfo{volume}{114}},
  \bibinfo{pages}{232301} (\bibinfo{year}{2015}).

\bibitem[{\citenamefont{Lorenz et~al.}(2015)\citenamefont{Lorenz, Hammer, and
  Mei\ss{}ner}}]{PhysRevD.92.034018}
\bibinfo{author}{\bibfnamefont{I.~T.} \bibnamefont{Lorenz}},
  \bibinfo{author}{\bibfnamefont{H.-W.} \bibnamefont{Hammer}},
  \bibnamefont{and} \bibinfo{author}{\bibfnamefont{U.-G.}
  \bibnamefont{Mei\ss{}ner}}, \bibinfo{journal}{Phys. Rev. D}
  \textbf{\bibinfo{volume}{92}}, \bibinfo{pages}{034018}
  (\bibinfo{year}{2015}).

\bibitem[{\citenamefont{Brodsky and de~Teramond}(2008)}]{Brodsky:2007hb}
\bibinfo{author}{\bibfnamefont{S.~J.} \bibnamefont{Brodsky}} \bibnamefont{and}
  \bibinfo{author}{\bibfnamefont{G.~F.} \bibnamefont{de~Teramond}},
  \bibinfo{journal}{Phys.Rev.} \textbf{\bibinfo{volume}{D77}},
  \bibinfo{pages}{056007} (\bibinfo{year}{2008}).

\bibitem[{\citenamefont{Matveev et~al.}(1973)\citenamefont{Matveev, Muradyan,
  and Tavkhelidze}}]{Matveev:1973uz}
\bibinfo{author}{\bibfnamefont{V.}~\bibnamefont{Matveev}},
  \bibinfo{author}{\bibfnamefont{R.}~\bibnamefont{Muradyan}}, \bibnamefont{and}
  \bibinfo{author}{\bibfnamefont{A.}~\bibnamefont{Tavkhelidze}},
  \bibinfo{journal}{Teor.Mat.Fiz.} \textbf{\bibinfo{volume}{15}},
  \bibinfo{pages}{332} (\bibinfo{year}{1973}).

\bibitem[{\citenamefont{Brodsky and Farrar}(1973)}]{Brodsky:1973kr}
\bibinfo{author}{\bibfnamefont{S.~J.} \bibnamefont{Brodsky}} \bibnamefont{and}
  \bibinfo{author}{\bibfnamefont{G.~R.} \bibnamefont{Farrar}},
  \bibinfo{journal}{Phys.Rev.Lett.} \textbf{\bibinfo{volume}{31}},
  \bibinfo{pages}{1153} (\bibinfo{year}{1973}).

\bibitem[{\citenamefont{Tomasi-Gustafsson and
  Rekalo}(2001)}]{TomasiGustafsson:2001za}
\bibinfo{author}{\bibfnamefont{E.}~\bibnamefont{Tomasi-Gustafsson}}
  \bibnamefont{and} \bibinfo{author}{\bibfnamefont{M.}~\bibnamefont{Rekalo}},
  \bibinfo{journal}{Phys.Lett.} \textbf{\bibinfo{volume}{B504}},
  \bibinfo{pages}{291} (\bibinfo{year}{2001}).

\bibitem[{\citenamefont{Ambrogiani et~al.}(1999)}]{Ambrogiani:1999bh}
\bibinfo{author}{\bibfnamefont{M.}~\bibnamefont{Ambrogiani}}
  \bibnamefont{et~al.} (\bibinfo{collaboration}{E835 Collaboration}),
  \bibinfo{journal}{Phys.Rev.} \textbf{\bibinfo{volume}{D60}},
  \bibinfo{pages}{032002} (\bibinfo{year}{1999}).

\bibitem[{\citenamefont{Hofstadter et~al.}(1960)\citenamefont{Hofstadter,
  Bumiller, and Croissiaux}}]{Hofstadter:1960zz}
\bibinfo{author}{\bibfnamefont{R.}~\bibnamefont{Hofstadter}},
  \bibinfo{author}{\bibfnamefont{F.}~\bibnamefont{Bumiller}}, \bibnamefont{and}
  \bibinfo{author}{\bibfnamefont{M.}~\bibnamefont{Croissiaux}},
  \bibinfo{journal}{Phys.Rev.Lett.} \textbf{\bibinfo{volume}{5}},
  \bibinfo{pages}{263} (\bibinfo{year}{1960}).

\bibitem[{\citenamefont{Yamada}(1971)}]{Yamada:1971ta}
\bibinfo{author}{\bibfnamefont{M.}~\bibnamefont{Yamada}},
  \bibinfo{journal}{Prog.Theor.Phys.} \textbf{\bibinfo{volume}{46}},
  \bibinfo{pages}{865} (\bibinfo{year}{1971}).

\bibitem[{\citenamefont{Shirkov and Solovtsov}(1997)}]{Shirkov:1997wi}
\bibinfo{author}{\bibfnamefont{D.}~\bibnamefont{Shirkov}} \bibnamefont{and}
  \bibinfo{author}{\bibfnamefont{I.}~\bibnamefont{Solovtsov}},
  \bibinfo{journal}{Phys.Rev.Lett.} \textbf{\bibinfo{volume}{79}},
  \bibinfo{pages}{1209} (\bibinfo{year}{1997}).

\bibitem[{\citenamefont{Kuraev}(2008)}]{Kuraev}
\bibinfo{author}{\bibfnamefont{E.~A.} \bibnamefont{Kuraev}},
  \bibinfo{journal}{private communication}  (\bibinfo{year}{2008}).

\bibitem[{\citenamefont{Bianconi and Radici}(1996)}]{Bianconi:1996ur}
\bibinfo{author}{\bibfnamefont{A.}~\bibnamefont{Bianconi}} \bibnamefont{and}
  \bibinfo{author}{\bibfnamefont{M.}~\bibnamefont{Radici}},
  \bibinfo{journal}{Phys.Rev.} \textbf{\bibinfo{volume}{C54}},
  \bibinfo{pages}{3117} (\bibinfo{year}{1996}).

\bibitem[{\citenamefont{Bijker and Iachello}(2004)}]{Bijker:2004yu}
\bibinfo{author}{\bibfnamefont{R.}~\bibnamefont{Bijker}} \bibnamefont{and}
  \bibinfo{author}{\bibfnamefont{F.}~\bibnamefont{Iachello}},
  \bibinfo{journal}{Phys.Rev.} \textbf{\bibinfo{volume}{C69}},
  \bibinfo{pages}{068201} (\bibinfo{year}{2004}).

\bibitem[{\citenamefont{Adamuscin et~al.}(2005)\citenamefont{Adamuscin,
  Dubnicka, Dubnickova, and Weisenpacher}}]{Adamuscin:2005aq}
\bibinfo{author}{\bibfnamefont{C.}~\bibnamefont{Adamuscin}},
  \bibinfo{author}{\bibfnamefont{S.}~\bibnamefont{Dubnicka}},
  \bibinfo{author}{\bibfnamefont{A.}~\bibnamefont{Dubnickova}},
  \bibnamefont{and}
  \bibinfo{author}{\bibfnamefont{P.}~\bibnamefont{Weisenpacher}},
  \bibinfo{journal}{Prog.Part.Nucl.Phys.} \textbf{\bibinfo{volume}{55}},
  \bibinfo{pages}{228} (\bibinfo{year}{2005}).

\bibitem[{\citenamefont{Belushkin et~al.}(2007)\citenamefont{Belushkin, Hammer,
  and Meissner}}]{Belushkin:2006qa}
\bibinfo{author}{\bibfnamefont{M.}~\bibnamefont{Belushkin}},
  \bibinfo{author}{\bibfnamefont{H.-W.} \bibnamefont{Hammer}},
  \bibnamefont{and} \bibinfo{author}{\bibfnamefont{U.-G.}
  \bibnamefont{Meissner}}, \bibinfo{journal}{Phys.Rev.}
  \textbf{\bibinfo{volume}{C75}}, \bibinfo{pages}{035202}
  (\bibinfo{year}{2007}).

\bibitem[{\citenamefont{Lomon and Pacetti}(2012)}]{Lomon:2012pn}
\bibinfo{author}{\bibfnamefont{E.~L.} \bibnamefont{Lomon}} \bibnamefont{and}
  \bibinfo{author}{\bibfnamefont{S.}~\bibnamefont{Pacetti}},
  \bibinfo{journal}{Phys.Rev.} \textbf{\bibinfo{volume}{D85}},
  \bibinfo{pages}{113004} (\bibinfo{year}{2012}).

\bibitem[{\citenamefont{de~Melo et~al.}(2004)\citenamefont{de~Melo, Frederico,
  Pace, and Salme}}]{deMelo:2003uk}
\bibinfo{author}{\bibfnamefont{J.}~\bibnamefont{de~Melo}},
  \bibinfo{author}{\bibfnamefont{T.}~\bibnamefont{Frederico}},
  \bibinfo{author}{\bibfnamefont{E.}~\bibnamefont{Pace}}, \bibnamefont{and}
  \bibinfo{author}{\bibfnamefont{G.}~\bibnamefont{Salme}},
  \bibinfo{journal}{Phys.Lett.} \textbf{\bibinfo{volume}{B581}},
  \bibinfo{pages}{75} (\bibinfo{year}{2004}).

\bibitem[{\citenamefont{Kuraev et~al.}(2012)\citenamefont{Kuraev,
  Tomasi-Gustafsson, and Dbeyssi}}]{Kuraev:2011vq}
\bibinfo{author}{\bibfnamefont{E.}~\bibnamefont{Kuraev}},
  \bibinfo{author}{\bibfnamefont{E.}~\bibnamefont{Tomasi-Gustafsson}},
  \bibnamefont{and} \bibinfo{author}{\bibfnamefont{A.}~\bibnamefont{Dbeyssi}},
  \bibinfo{journal}{Phys.Lett.} \textbf{\bibinfo{volume}{B712}},
  \bibinfo{pages}{240} (\bibinfo{year}{2012}).

\end{thebibliography}
\end{document}